\documentclass[aps,twocolumn,amsmath,amssymb,preprintnumbers,floatfix,prb,superscriptaddress,longbibliography]{revtex4-2}

\usepackage{comment}
\usepackage[version=4]{mhchem}
\usepackage[utf8]{inputenc}
\usepackage{newtxtext}
\usepackage[upint]{newtxmath}
\usepackage{microtype}
\usepackage{textcomp}
\usepackage{dsfont}
\usepackage{eucal}
\usepackage{siunitx}
\usepackage{soul}

\usepackage{enumerate}
\usepackage{amsfonts}
\usepackage{color}
\usepackage{soul}

\usepackage{todonotes}
\presetkeys%
    {todonotes}%
    {inline}{}

\usepackage{graphicx}

\newcommand{\ca}[2][]{c_{#2}^{\vphantom{\dagger}#1}} 
\newcommand{\cc}[2][]{c_{#2}^{{\dagger}#1}}                                    

\newcommand{\vecP}{\boldsymbol{P}}
\newcommand{\vecM}{\boldsymbol{M}}

\newcommand{\vk}{\boldsymbol{k}}

\renewcommand{\v}[1]{\boldsymbol{#1}}                           
       
\definecolor{darkgreen}{rgb}{0.00,0.6,0.00}
\definecolor{DarkBlue}{rgb}{0,0,0.80}
\definecolor{DarkRed}{rgb}{0.80,0,0}
\definecolor{Purple}{rgb}{0.55,0,0.55}
\definecolor{Purple}{rgb}{0,0,0.8}

\newcommand{\revision}[1]{\textcolor{black}{#1}}

\usepackage[colorlinks=true,urlcolor=blue,citecolor=blue,linkcolor=blue]{hyperref}

\DeclareMathOperator*{\avg}{avg} 

\newcommand{\eg}{e.g.\ }

\let\epsilon\varepsilon

\begin{document}

\title{The fate of $p$-wave spin polarization in helimagnets with Rashba spin-orbit coupling}
\author{Erik W. Hodt}
\affiliation{Center for Quantum Spintronics, Department of Physics, Norwegian \\ University of Science and Technology, NO-7491 Trondheim, Norway}
\author{Hendrik Bentmann}
\affiliation{Center for Quantum Spintronics, Department of Physics, Norwegian \\ University of Science and Technology, NO-7491 Trondheim, Norway}
\author{Jacob Linder}
\affiliation{Center for Quantum Spintronics, Department of Physics, Norwegian \\ University of Science and Technology, NO-7491 Trondheim, Norway}\date{\today}
\begin{abstract}
It has recently been realized that magnetic systems with coplanar magnetic order that are invariant under the combined operation of time-reversal and translation with half a unit cell feature energy bands  with a symmetry-protected \textit{p}-wave spin polarization. Such $p$-wave magnets are a sought-after spin analogy of unconventional triplet superconducting pairing and show promise for use in spintronics. Metallic helimagnets provide a realization of $p$-wave magnetism, but such order often occurs in systems lacking inversion symmetry so that Rashba spin-orbit interactions can be prominent. An important question is therefore how the magnitude and the existence of $p$-wave spin polarization is affected by Rashba spin-orbit interaction. Here, we prove that while the $p$-wave symmetry of the spin-polarized bands is strictly speaking removed by \revision{such} spin-orbit interactions in helimagnets unless the period of the helix is fine-tuned, the actual quantitative deviation from $p$-wave symmetry is extremely weak unless the period of the helix is only a few lattice sites. Thereafter, we show that the $p$-wave magnetism becomes completely robust in pairs of antiferromagnetically coupled helices. More precisely, the $p$-wave spin-polarization of the bands then appears regardless of the periodicity and regardless of the strength of the spin-orbit interactions. This shows
that 
antiferromagnetically coupled
helimagnetic chains produce robust $p$-wave spin
polarization free of fine-tuning requirements, making them
attractive for potential spintronic applications.
\end{abstract}

\maketitle

\section{Introduction}
Magnetic materials form a centerpiece in the field of condensed matter physics. Besides their widespread use in applications, ranging from various electronic devices to medical technology, such materials reveal a wealth of interesting fundamental quantum physics. There is a zoo of magnetic states in materials, ranging from traditional ferromagnets and antiferromagnets to more complex magnetic configurations like skyrmions \cite{nagaosa_natnano_13, fert_natrevmat_17} and spin ice \cite{castelnovo_annrevcmp_12, skjaervo_natrevphys_20}.

The behavior of conduction electrons in a magnetic material is heavily influenced by the type of magnetism that is exhibited. This is a key principle in the field of spintronics \cite{hirohata_jmmm_20} where the aim is to utilize the electron spin for practical purposes rather than its electric charge. In this context, it is particularly useful to identify ways that enable the motion of the electron to be controlled via its spin, or vice versa. The realization that a large group of antiferromagnetic materials display such spin-momentum coupling \cite{noda_pccp_16, ahn_prb_19, hayami_jpsj_19, yuan_prb_20, smejkal_sciadv_20} and form a material class known as altermagnets \cite{smejkal_prx_22a, smejkal_prx_22b} has therefore generated much interest recently. In altermagnets, the electrons exhibit a momentum-dependent spin polarization while the material as a whole has no net magnetization, making them of high interest for spintronics \cite{bai_arxiv_24}. 

Interestingly, the precise form of the spin-momentum coupling in some altermagnets is a magnetic analogy of how electrons in high-temperature superconductors form pairs, and has a so-called $d$-wave symmetry \cite{keimer_nature_15}. Other types of symmetries, such as $p$-wave, can be realized in superfluid He-3 \cite{leggett_rmp_75}. It turns out that such a state also has a magnetic analogy. Helimagnetic systems where the localized spin moments display a spatially rotating pattern have been noted to endow electrons with a $p$-wave spin polarization \cite{martin_prb_12, chen_2dmater_22, mayo_arxiv_24, Hayami2020, Hayami2020_2, Hayami2020_3}, meaning that electrons with opposite momenta have opposite spin expectation values. The possibility to realize $p$-wave magnetic phases from spin-independent interactions, via a so-called Pomeranchuk instability has also been studied \cite{wu_prb_07, jung_prb_15, kiselev_prb_17, wu_prb_18}, but not experimentally realized so far. Very recently, a symmetry-analysis of the Hamiltonian for a helimagnetic system revealed under which circumstances $p$-wave spin polarization appears for the itinerant electrons \cite{kudasov_prb_24}, and the importance of $\hat{\mathcal{T}}\hat{\boldsymbol{\tau}}$-symmetry (where $\hat{\mathcal{T}}$ is the time-reversal operation and $\hat{\boldsymbol{\tau}}$ is a lattice translation operation) was highlighted in \cite{hellenes_arxiv_23} and used to propose that CeNiAsO is a $p$-wave magnet. A minimal effective model capturing the main qualitative features of $p$-wave magnets, permitting an analytical approach, have recently been developed and used to predict large tunneling magnetoresistance (TMR) values \cite{brekke_prl_24}. 

$p$-wave magnets look promising for spintronics applications due to the strong (non-relativistic) spin-momentum coupling.  \revision{Promising predictions include strong TMR values, spin-anisotropic transport and the possibility of charge-to-spin conversion \cite{brekke_prl_24}, a non-relativistic Edelstein effect \cite{edelstein_pwave_2, pwave_edelstein} and the presence of a spin-galvanic effect \cite{bergeret_pwave}.} The extent to which the characteristic $p$-wave symmetry can be utilized in realistic devices depends crucially on how robust the $p$-wave spin-polarization stemming from the magnetic texture is against other interactions that may be present in the material. Of particular importance is how \revision{Rashba-type} spin-orbit interactions, which will be especially prominent in crystals lacking inversion symmetry or in thin films, affect the magnitude and existence of the $p$-wave spin polarization of the electrons stemming from the magnetic texture. The answer to this question will be decisive in determining whether or not $p$-wave magnetism in for instance helimagnets is a robust feature that occurs irrespective of the details of the helimagnetic pattern and irrespective of the presence of spin-orbit interactions stemming from the lack of structural or bulk inversion symmetry. \revision{While a breaking of bulk inversion symmetry could in principle lead to a Dresselhaus-type spin-orbit interaction, we focus on the Rashba-type in this manuscript. In low-dimensional devices, geometry and the presence of interfaces will generally provide sources of structural inversion symmetry breaking which through the emergence of Rashba-type spin-orbit interactions can interfere with the \textit{p}-wave properties.}

We here address \revision{the question of how Rashba-type SOC and \textit{p}-wave spin polarization interferes} and report two main results. The first result is that a \revision{Rashba-type} spin-orbit interaction induce a even-odd effect in a helimagnetic chain. With SOC, the composite time-reversal symmetry in helices with odd unit cells is lifted and a nonzero magnetization emerges. While SOC ruins the $p$-wave polarization for helimagnets with an odd number of spins in the period of their magnetic texture \cite{kudasov_prb_24}, it survives and remains even for strong spin-orbit interactions when the period has an even number of spins. While this suggests that helimagnets may not represent a robust realization of $p$-wave magnetism, since the precise number of sites in the period will be very difficult to control experimentally, our second main result overcomes this problem. The second result shows that when pairs of helimagnetic chains couple antiferromagnetically to each other, as experimentally realized \cite{bode_nature_07} in manganese atoms on tungsten (110), perfect $p$-wave polarization appears  regardless of the periodicity of the helimagnetic chain and regardless of the strength of the spin-orbit interactions. These results show that coupled helimagnetic chains can be used to realize robust $p$-wave spin polarization free of fine-tuning requirements, making them attractive for potential applications.

\section{Theory}
\begin{figure}
    \centering
    \includegraphics[width=0.99\linewidth]{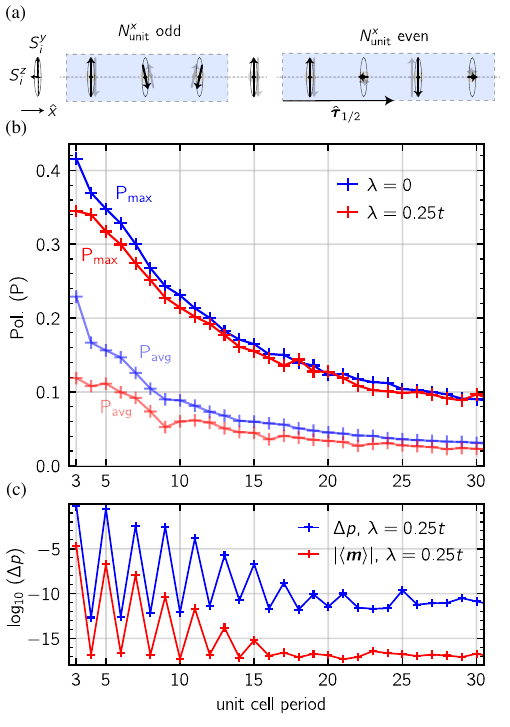}
    \caption{(a) In a single helimagnetic chain with odd unit cell length, there exists no lattice translation $\hat{\boldsymbol{\tau}}$ which maps the chain onto its time-reversed counterpart (grey arrows). On the contrary, for even unit cells, the combined operation of time-reversal $\hat{\mathcal{T}}$ and $\hat{\boldsymbol{\tau}}=\hat{\boldsymbol{\tau}}_{1/2}$ where $\hat{\boldsymbol{\tau}}_{1/2}$ is a translation by half a unit cell length, leaves the system invariant. (b) The maximum bandstructure spin-polarization (\textit{P}) is shown for a helimagnet with FM order between chains and $N_\text{u.c.}^x=3-30$ sites. The transparent lines show the average instead of the maximum \revision{spin} polarization [see Eq. (\ref{eqn: measure})]. Upon introduction of SOC ($\lambda=0.25t$), the \textit{x}-polarization of the bandstructure decreases as the odd in $k_y$ $\sigma_x$ from the SOC counteracts the $k_x \sigma_x$ of the helical structure. (c) The presence of SOC disrupts the perfect antisymmetric spin polarization for odd unit-cells which also manifests in an effective time-reversal breaking and the appearance of a non-zero $\langle m_{x/y} \rangle$ magnetization. For unit cells where $N_\text{u.c.}^x$ is even, the \textit{p}-wave order is protected by the $\hat{\mathcal{T}}\hat{\boldsymbol{\tau}}_{1/2}$ symmetry and is not lifted by the presence of SOC. $N_\text{Fourier}=50$ in both \textit{x} and \textit{y}.}
    \label{fig: FM}
\end{figure}

We model a 2D helimagnet with a tight-binding model where the itinerant electrons are coupled to localized and spatially varying moments through an \textit{s-d} coupling. The moments are taken to be collinearly ordered (AFM or FM) between chains in the \textit{y} direction while moments rotate in a non-collinear fashion along the chains in \textit{x}, giving rise to the helical magnetic texture. The case of a single helimagnetic chain is shown in Fig. \ref{fig: FM}(a). The Hamiltonian is given by

\begin{figure}
    \centering
    \includegraphics[width=0.99\linewidth]{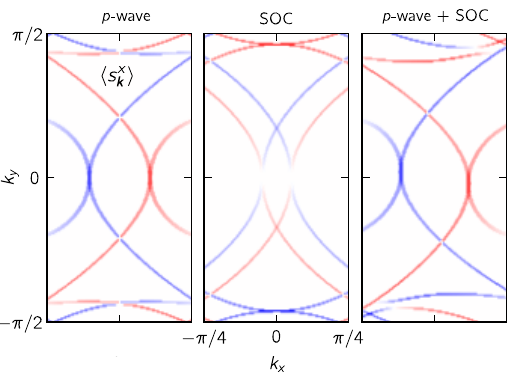}
    \caption{Fermi surface at $\mu=-2t$ for (a) a $N_\text{u.c.}^x=4$ helical structure $J_\text{sd}=t$ (b) Rashba metal ($\lambda=0.25t$) and (c) helical structure + Rashba SOC ($J_\text{sd}=t$, $\lambda=0.25t$). In the bare helical structure, the transverse $\langle s_{\boldsymbol{k}}^x\rangle$ polarization is asymmetrical under inversion of $k_x$, $(k_x, k_y)\rightarrow(-k_x, k_y)$. In absence of magnetic texture with Rashba SOC, the same \revision{spin} polarization is asymmetric with respect to the $k_y$ component, $(k_x, k_y)\rightarrow(k_x, -k_y)$. Combining the helical magnet with Rashba SOC renders the $\langle s_{\boldsymbol{k}}^x\rangle$ asymmetric only under inversion of the total momentum, $(k_x, k_y)\rightarrow(-k_x, -k_y)$. }
    \label{fig: FS}
\end{figure}
\begin{multline}
    H=-t\sum_{\langle i,j\rangle,\sigma} \cc[]{i,\sigma}\ca[]{j,\sigma} -J_\text{sd}\sum_{i,\sigma,\sigma'} (\boldsymbol{S}_i\cdot\boldsymbol{\sigma})_{\sigma\sigma'}^{\vphantom{\dagger}}\cc[]{i,\sigma}\ca[]{i,\sigma'} \\ -\frac{i\lambda}{2}\sum_{\langle i,j \rangle, \sigma,\sigma'}\cc[]{i,\sigma}(\boldsymbol{\sigma}_y \delta_x - \boldsymbol{\sigma}_x \delta_y)_{\sigma\sigma'}^{\vphantom{\dagger}}\ca[]{j,\sigma'}\label{eqn: H}
\end{multline}
where 
\begin{align}
\boldsymbol{S}_i = e^{i\eta \pi y_i}[\cos(\varphi x_i)\hat{\boldsymbol{y}} + \sin(\varphi x_i)\hat{\boldsymbol{z}}] \label{eqn:S}
\end{align}
and where $\eta=0(1)$ describe FM (AFM) order between the helically ordered chains. Above, $J_{\text{sd}}$ is the \textit{s-d} coupling strength, \textit{t} the regular tight-binding hopping parameter and $\lambda$ the strength of a Rashba spin-orbit coupling due to inversion-symmetry breaking in \textit{z}. The magnetic texture is rotating in the \textit{yz}-plane as a function of the \textit{x}-coordinate. \revision{We point out that we here assume the existence of a helimagnetic ground state through choosing an explicit form of $\boldsymbol{S}_i$. A more rigorous approach would be to obtain a helimagnetic ground state self-consistently through an appropriate mean-field decomposition of e.g. the Hubbard model or the $t-J$ model with appropriate parameters. As the focus of this article is on the effect of such a ground state on the itinerant electrons, we use the implementation in Eq. (\ref{eqn:S}).} 

The period over which the texture completes a full rotation naturally defines a unit cell with $N_{\text{u.c.}}^x$ sites and the relative angle between adjacent moments in \textit{x} is $\varphi= \frac{2\pi}{N_\text{u.c.}^x + 1}$. Assuming periodic boundary conditions in both \textit{x} and \textit{y}, we introduce Fourier-transformed operators with momenta running between 
\begin{gather}
k_x\in\big[-\frac{\pi}{N_{\text{u.c.}}^x}, \frac{\pi}{N_{\text{u.c.}}^x}\big),\\[5pt] k_y \in\big[-\frac{\pi}{2}, \frac{\pi}{2}\big)
\end{gather}
where we have set the unit cell dimension in \textit{y} to 2 to allow for AFM coupling between chains. In Fourier space, the Hamiltonian can be written as 
\begin{equation}
    H=\sum_{\boldsymbol{k}}\Tilde{c}_{\boldsymbol{k}}^\dagger h_{\boldsymbol{k}} \Tilde{c}_{\boldsymbol{k}}^{\vphantom{\dagger}} \label{eqn: H_fourier}
\end{equation}
where the Hamiltonian block $h_{\boldsymbol{k}}$ is $4N_\text{u.c.}^x\times 4N_\text{u.c.}^x $ (recall that $N_\text{u.c.}^y=2$) and where \begin{equation}\Tilde{c}_{\boldsymbol{k}}^{\vphantom{\dagger}}=(\ca[]{\boldsymbol{k}, 1, \uparrow}, ...,\ca[]{\boldsymbol{k}, \alpha, \uparrow}, ..., \ca[]{\boldsymbol{k}, 2N_\text{u.c.}^x, \downarrow})^T\end{equation} with $\alpha$ denoting the site index within a unit cell.

A given $\boldsymbol{k}$-block in Eq. ($\ref{eqn: H_fourier}$) is diagonalized by the unitary matrix $U_{\boldsymbol{k}}$ such that $h_{\boldsymbol{k}}^{\vphantom{\dagger}} =U_{\boldsymbol{k}}^\dagger \Lambda_{\boldsymbol{k}}^{\vphantom{\dagger}} U_{\boldsymbol{k}}^{\vphantom{\dagger}} $ where $\Lambda_{\boldsymbol{k}}^{\vphantom{\dagger}} =\text{diag}[E_{\boldsymbol{k}, 1} ... E_{\boldsymbol{k}, 4N_\text{u.c.}^x}]$ and 
$U_{\boldsymbol{k}}=[\Phi_{\boldsymbol{k}, 1} ... \Phi_{\boldsymbol{k}, 4N_\text{u.c.}^x}]$
with the eigenvectors 
\begin{align}
\Phi_{\boldsymbol{k},n}= [u_{\boldsymbol{k}, 1, n}\text{ }v_{\boldsymbol{k}, 1, n} \text{ }...\text{ } v_{\boldsymbol{k}, 2N_\text{u.c.}^x, n}  ]^T.
\end{align}

The magnetic texture rotates in the \textit{yz}-plane, and in the following, the transverse \textit{x}-polarization of itinerant electrons will be of particular interest. Expressing the original fermion operators in the new basis, it follows that the \textit{x}-polarization of the momentum mode $\boldsymbol{k}$ can be expressed as 
\begin{align}
    \langle S_{\boldsymbol{k}}^x \rangle &= \frac{1}{2}\sum_{\alpha=1}^{2N_\text{u.c.}^x}\sum_{n = 1}^{4N_\text{u.c.}^x}\big(u_{\boldsymbol{k}, \alpha, n}^* v_{\boldsymbol{k}, \alpha, n} \notag+ u_{\boldsymbol{k}, \alpha, n}v_{\boldsymbol{k}, \alpha, n}^*\big)n_F(E_{\boldsymbol{k},n}) \notag\\
    &= \sum_{n=1}^{4N_\text{u.c.}^x}\langle s_{n, \boldsymbol{k}}^x \rangle n_F(E_{\boldsymbol{k},n})
\end{align}
where $\alpha$ denotes the particular site in the unit cell and where \textit{n} is the band index. We have introduced the band-resolved \revision{spin} polarization 
\begin{equation}
\langle s_{n, \boldsymbol{k}}^x \rangle=\frac{1}{2}\sum_{\alpha=1}^{2N_\text{u.c.}^x}u_{\boldsymbol{k}, \alpha, n}^* v_{\boldsymbol{k}, \alpha, n} + u_{\boldsymbol{k}, \alpha, n}v_{\boldsymbol{k}, \alpha, n}^* \label{band-resolved polarization}
\end{equation}
which gives the \revision{spin} polarization of a state in a particular band \textit{n} at a particular $\boldsymbol{k}$ in the Brillouin zone. We can likewise define the band-resolved \textit{y} and \textit{z} polarizations,
\begin{align}
    \langle s_{n, \boldsymbol{k}}^y \rangle&=-\frac{i}{2}\sum_{\alpha=1}^{2N_\text{u.c.}^x}u_{\boldsymbol{k}, \alpha, n}^* v_{\boldsymbol{k}, \alpha, n} - u_{\boldsymbol{k}, \alpha, n} v_{\boldsymbol{k}, \alpha, n}^*  \\
    \langle s_{n, \boldsymbol{k}}^z \rangle&=\frac{1}{2}\sum_{\alpha=1}^{2N_\text{u.c.}^x}|u_{\boldsymbol{k}, \alpha, n}|^2 - |v_{\boldsymbol{k}, \alpha, n}|^2 \label{band-resolved xy polarization},
\end{align}
but these are zero in the absence of SOC due to the compensated nature of the helical unit cell. \revision{Note that in the following, any reference to spin-orbit coupling imply a SOC of Rashba type.} 

A defining characteristic of \textit{p}-wave magnetism is the emergence of a non-zero transverse  \revision{spin} polarization of the electron bands, $\langle s_{n, \boldsymbol{k}}^x \rangle \neq 0$, despite the moments of $\boldsymbol{S}_i$ of Eq. (\ref{eqn: H}) rotating in the \textit{yz}-plane. For a given wavevector $\boldsymbol{k}$, this polarization is odd under spatial inversion $\langle s_{n, \boldsymbol{k}}^x \rangle = -\langle s_{n, -\boldsymbol{k}}^x \rangle$ such that the overall system is compensated. Rashba spin-orbit coupling also induces an antisymmetric spin-polarization. We plot the Fermi surface in Fig. \ref{fig: FS} at $\mu=-2t$ for a \textit{p}-wave magnet without SOC (a), for a Rashba metal with $J_\text{sd}=0$ (b) and for a combined \textit{p}-wave + SOC system in (c). The color denotes the \revision{spin} polarization in \textit{x} of the bands crossing the Fermi level. We note that the $\langle s_{n, \boldsymbol{k}}^x \rangle$ induced by the helical texture is odd in $k_x$, the direction along which the texture rotates. Likewise, $\langle s_{n, \boldsymbol{k}}^x \rangle$ is odd in $k_y$ for SOC, due to the term $\sigma_x k_y$ in the Hamiltonian. While both are antisymmetric with respect to a full spatial inversion, the particular momentum component giving rise to the inversion is unique. When \textit{p}-wave and SOC is combined, the result can be a \revision{spin} polarization $\langle s_{n, \boldsymbol{k}}^x \rangle$ whose axis of inversion is tilted away from the \textit{x}-axis ($k_x\rightarrow -k_x$) in the bare \textit{p}-wave magnet, but we emphasize that it is not obvious that the perfectly antisymmetric spin polarization under $\boldsymbol{k}\rightarrow-\boldsymbol{k}$ is preserved for systems with both a helimagnetic texture and SOC present. In fact, for a helimagnetic chain with an odd unit cell period, the antisymmetric spin polarization is absent in the presence of Rashba SOC. We also point out that while Rashba SOC induces a $\sigma_y$ component odd in $k_x$, we shall ignore this component in the following as it is not related to the \textit{p}-wave magnet.

A key question in this paper is how the \textit{p}-wave spin polarization stemming from the magnetism is affected when \revision{Rashba-type} SOC is introduced. To answer this, we use two measures originating from the above characteristics from which the \textit{p}-wave order can be ascertained. (1) In order for the system to depict \textit{p}-wave order, the maximum \revision{spin} polarization of the bandstructure must be non-zero. However, recalling that Rashba SOC induces an \textit{x}-polarization on its own which is odd in $k_y$, a measure is needed that can distinguish between contributions from the texture and from SOC. We argue that as the texture-induced \revision{spin} polarization is odd in $k_x$, we obtain the ``effective" \textit{p}-wave polarization induced by the magnetic texture by summing the band-resolved \textit{x}-polarization $\langle s_{n,\boldsymbol{k}}^x \rangle$ over all $k_y$ before taking either the maximum magnitude or average magnitude. We then expect the odd contribution from SOC to cancel, serving as an indicator of the remaining \textit{p}-wave polarization,
\begin{equation}
P=\max_{n,k_x} \bigg|\sum_{k_y} \langle {s}_{n, k}^x \rangle \bigg| \; \bigg/ \; \avg_{n,k_x} \bigg| \sum_{k_y} \langle {s}_{n, k}^x \rangle \bigg|\label{eqn: measure}
\end{equation}
We point out that for a system with only Rashba SOC, this measure gives zero \textit{p}-wave polarization. \

\revision{The mere existence of a spin polarized bandstructure is not sufficient for a system to depict \textit{p}-wave character.}  For a \textit{p}-wave magnet, this nonzero \revision{spin} polarization must be odd under a spatial inversion. We therefore define the deviation from \textit{p}-wave order $\Delta p$,
\begin{equation}
    \Delta p = \max_{n,\boldsymbol{k}}\big\{\big|\langle \boldsymbol{s}_{n, \boldsymbol{k}} \rangle + \langle \boldsymbol{s}_{n,- \boldsymbol{k}} \rangle \rangle \big|\big\}
\end{equation}
where we have defined the vector $\langle \boldsymbol{s}_{n,\boldsymbol{k}}\rangle = (\langle s_{n,\boldsymbol{k}}^x\rangle, \langle s_{n,\boldsymbol{k}}^y\rangle, \langle s_{n,\boldsymbol{k}}^z\rangle)$. In the following, we shall take a system where  $P\neq0$ and $\Delta p = 0$ to be a \textit{p}-wave magnet.

To obtain an intuition behind the emergence of the \revision{spin}-polarization in \textit{x} as well as the dependence on the unit cell length, we briefly lay aside the lattice model and consider a free electron coupled to a rotating magnetic field through the coupling strength $\gamma$,
\begin{equation}
    H=-\frac{1}{2m}\nabla^2 - \gamma \boldsymbol{\sigma}\cdot \boldsymbol{B}(x,y)
\end{equation}
where the magnetic field is given by 
\begin{equation}
\boldsymbol{B}(x,y)=B_0[\cos(\varphi_x x_i)\hat{\boldsymbol{y}} + \sin(\varphi_y x_i)\hat{\boldsymbol{z}}]
\end{equation}
 and where we have allowed for a spatially dependent texture in both \textit{x} and \textit{y} described by the spatial frequencies $\varphi_x$ and $\varphi_y$ respectively. \revision{This leads to the following eigenvalueproblem,
\begin{multline}
    E\psi(x,y)=\bigg\{-\frac{1}{2m}\nabla^2\mathbb{I} \\-\gamma B_0
    \begin{bmatrix}
        \sin(\varphi_xx+\varphi_yy) & -i\cos(\varphi_xx+\varphi_yy) \\
        i\cos(\varphi_xx+\varphi_yy)& -\sin(\varphi_xx+\varphi_yy)
    \end{bmatrix} 
    \bigg\} \psi(x,y) \label{eqn: 4}
\end{multline}
where $\psi(x,y)=\begin{pmatrix}
    \psi_1 & \psi_2
\end{pmatrix}^T$ is a spinor wavefunction.} We can reinstate translational invariance with the unitary transformation $\mathcal{R}=e^{-i\sigma_x(\varphi_xx+\varphi_yy)/2}$ which introduces a spin reference frame rotating around the \textit{x}-axis. \revision{Details from a similar derivation can be found in Ref. \cite{Calvo}}. The eigenfunctions are given by
\begin{equation}
    \psi_\pm(x,y)=e^{i\boldsymbol{k}\boldsymbol{r}}e^{-i\sigma_x(\varphi_xx+\varphi_yy)/2}e^{-i\sigma_z\phi(\boldsymbol{k})/2}\eta_{\pm}
\end{equation}
where $\eta_\pm=(1,0)^T/(0,1)^T$ and where 
\begin{equation}
\phi(\boldsymbol{k})=\text{atan2}\big[-A,-(\varphi_xk_x+\varphi_yk_y)\big]
\end{equation}
The spin polarization of the eigenfunctions is given by
\begin{align}
\langle S^x\rangle_{\pm}(x,y) &= \mp \cos(\phi(\boldsymbol{k})) \label{eqn: 11} \\
    \langle S^y\rangle_{\pm}(x,y) &=\mp \sin(\phi(\boldsymbol{k}))\cos(\varphi_xx+\varphi_yy) \\
    \langle S^z\rangle_{\pm}(x,y) &=\mp \sin(\phi(\boldsymbol{k}))\sin(\varphi_xx+\varphi_yy) \label{eqn: 13}
\end{align}

 A key observation is now that while the expectation value of the \textit{y} and \textit{z} \revision{spin} polarization rotates in space as the electron propagates through the rotating magnetic field, a transverse \textit{x} polarization also arises when $\boldsymbol{\varphi}\cdot\boldsymbol{k}$ becomes large due to the rotating spin-reference frame. The strength of this polarization scales inversely with the period $\lambda$ of the helix ($\varphi_{x/y} = 2\pi/\lambda_{x/y}$) so that helices with a fast rotation induces a stronger transverse \revision{spin} polarization for a given electron wavevector $\boldsymbol{k}$. We can thus make two predictions which will be important in the following: 1) For a system with helical chains with collinear spins in \textit{y}, $\varphi_y=0$ and $\langle S_x \rangle$ should vanish as the length of the unit cell increases ($\varphi_x\rightarrow 0$). 2) In the presence of a spatial dependence in the texture along \textit{y} ($\varphi_y \neq 0$), for instance realized through an antiferromagnetic coupling along \textit{y}, we predict a contribution to the \textit{x}-polarization which is independent of $N_\text{u.c.}^x$ and thus remains even as $\varphi_x\rightarrow0$. 

We end this section by elaborating on the differences between the polarization induced by the texture and Rashba SOC. An important point is that Rashba SOC can on its own induce a spin polarization in the electron bands that is odd under inversion of momentum, $\vk \to -\vk$. However, in addition to having a spin expectation value that is not locked to one particular quantization axis as one moves around the Fermi surface, there are several qualitative differences from the $p$-wave spin polarization induced by magnetic textures. Firstly, the origin of Rashba SOC is relativistic, generally making it a much smaller effect than the $p$-wave magnetism originating from the exchange interaction. Secondly, Rashba SOC will appear due to structural inversion asymmetry breaking, such as close to interfaces or edges. For thin-films, Rashba SOC can thus be influential in the entire system. In contrast, magnetic textures arising from magnetic frustration do not require inversion symmetry breaking and are not confined to surfaces or edges, allowing it as a bulk effect. Thirdly, the $p$-wave spin polarization induced by certain magnetic textures is protected by a $\hat{\mathcal{T}}\hat{\boldsymbol{\tau}}$-symmetry. Chains with an odd number of sites in their magnetic period do not have this symmetry, and, as originally discussed in Ref. \cite{kudasov_prb_24} and quantified in the present manuscript, any field breaking spin-rotational invariance (such as SOC) will ruin the $p$-wave polarization coming from the magnetic texture \cite{kudasov_prb_24}. Therefore, although SOC and magnetic textures on their own can induce antisymmetric ($p$-wave) spin polarization, their combination, which can appear \eg in thin-films, can destroy the $p$-wave polarization of the itinerant electrons. 

Therefore, in order to utilize the potentially large $p$-wave polarization coming from helimagnets in spintronics, it is necessary to determine under which circumstances it exists despite the presence of SOC, and moreover how the magnitude of the $p$-wave polarization coming from the helimagnet is affected by SOC. These are the questions we answer in this manuscript.

\section{Results and discussion}
In the case of FM order between the helical chains, we observe an even-odd effect in $\Delta p$ when \revision{Rashba} SOC is introduced, but the effect of SOC on bandstructure polarization itself is relatively weak. The \revision{spin} polarization \textit{P} and deviation from \textit{p}-wave order $\Delta p$ are shown in Fig. \ref{fig: FM}(b)-(c) for unit cells with lengths ranging from $N_\text{u.c.}^x=3-30$ lattice sites with SOC (red line, $\lambda=0.25t$) and without SOC (blue line). The number of Fourier modes is taken to be $N_\text{Fourier}=50$ in both \textit{x} and \textit{y}. In the absence of SOC, a \textit{p}-wave $x$-polarization emerges which remains independently of the odd/even nature of the unit cell and which declines in magnitude as the period of the magnetic texture becomes larger. The emergence of the transversal \revision{spin} polarization is not inherently related to the even/odd nature of the unit cell and its independence of the particular unit cell realization in absence of SOC is thus to be expected. $\Delta p$ is zero for all periods considered in the absence of SOC and is not plotted. The decline of the \revision{spin} polarization as function of unit cell length can be understood in two ways: 1) The continuum model predicts that the magnitude of the transverse polarization goes like $\boldsymbol{\varphi}\cdot\boldsymbol{k}$. As the period of the texture becomes longer, $\boldsymbol{\varphi}$ decreases and thus also the magnitude of the \revision{spin} polarization. 2) In the limit of an infinite unit cell, our system should behave as a ferromagnet polarized in \textit{y}. With spin-split bands, a polarization in \textit{x} is not permitted and it thus has to vanish as the system gradually becomes more ferromagnetic like. 

By introducing a relatively strong SOC, we observe a small change in the polarization measure itself (\textit{P}), but the primary effect of SOC is not to disrupt the \textit{x}-polarization of individual states. Instead, the SOC leaves a signature in the emergence of a \textit{p}-wave symmetry lifting when the number of spins in the unit cell is odd, indicated by $\Delta p\neq0$. This symmetry-breaking is also associated with a breaking of the composite time-reversal symmetry made visible through the emergence of a nonzero $\langle m_{x/y}\rangle$ magnetization. Note that $\Delta p$ and $\langle |\boldsymbol{m}|\rangle$ are plotted on a logarithmic scale and that both fall of quickly with unit cell length. As such, the SOC-induced lifting of \textit{p}-wave symmetry might be difficult to detect, even for relatively small unit cell periods of 5-10 sites.  \revision{We here emphasize that the essential observable indicating the lifting of the \textit{p}-wave state is the breaking of the perfect antisymmetric property of the spin polarization. While the introduction of SOC is not detrimental to the existence of spin polarization in the bandstructure (Fig. \ref{fig: FM}(b)), this is to be expected becaused both the \textit{p}-wave state and Rashba SOC gives rise to a bandstructure spin polarization. The crucial detail here is that in a system combining a helimagnet and Rashba SOC, the perfect antisymmetric property of this spin polarization, the \textit{p}-wave nature, is inherently present or absent depending on the size of the unit cell. The fact that spin-rotation symmetry is lifted in systems with Rashba SOC and the subsequent emergence of a non-zero magnetization for textures with odd unit cells may pose a challenge for applications relying explicitly on a the use of an exact \textit{p}-wave symmetry of the spin-polarization and absence of net magnetization.}

The even-odd effect can intuitively be understood by recalling that for odd unit cells, the \textit{p}-wave order is not protected by a $\hat{\mathcal{T}}\hat{\boldsymbol{\tau}}_{1/2}$ operation, time-reversal combined with a lattice translation, as no appropriate translation exists which maps the time-reversed lattice back onto itself. In the absence of SOC, however, the invariance of the lattice is still maintained as a $\hat{C}_2$ rotation in spin-space is a symmetry operation of the system. This symmetry is broken when SOC is introduced and the spin-rotation symmetry lifted. 

\begin{figure}
    \centering
    \includegraphics[width=0.99\linewidth]{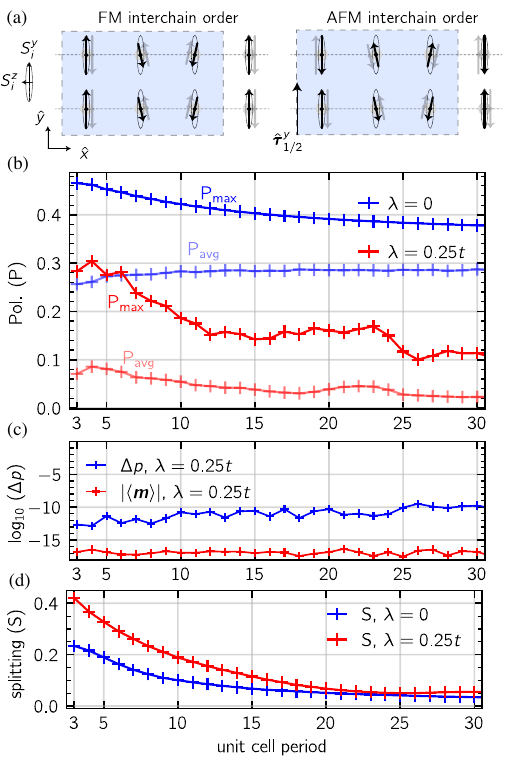}
    \caption{(a) Restricting the \textit{y}-order to collinear, the chains can be ordered in a FM or AFM fashion pertaining to the relative orientation of moments in the two chains. For FM interchain order, the nature of the \textit{p}-wave order is dictated by the presence (even) or absence (odd) of a translation vector mapping the time-reversed lattice (grey arrows) onto itself. For AFM interchain order, this translation vector $\hat{\boldsymbol{\tau}}_{1/2}$ is present independent of unit cell length as the AFM order makes the two rows of the unit cell related by the time-reversal operation. (b) Maximum of the \textit{p}-wave bandstructure \revision{spin} polarization (P)  is shown for a system with AFM interchain coupling and $N_\text{u.c.}^x=3-30$. Transparent lines show average instead of maximum. The effect of SOC on the \revision{spin} polarization is more significant than in the FM case, and the polarization in absence of SOC is particularly stable as function of unit cell period. (c) Deviation from antisymmetric spin polarization and magnitude of magnetization is shown. The guaranteed presence of a translation vector $\hat{\boldsymbol{\tau}}_{1/2}$ destroys the even-odd effect from the FM case Fig. \ref{fig: FM} indicating a stable \textit{p}-wave order independent on the magnetic texture period. (d) As the unit cell becomes longer, the system approaches an A-type antiferromagnet which should have spin-degenerate bands. While a nonzero $\langle s_{n,\boldsymbol{k}}^x\rangle$ is allowed in the AFM due to the degenerate bands, such a band-resolved \revision{spin} polarization is not useful since the bands are spin-degenerate in a conventional AFM. We therefore plot the averaged level splitting $S$ between energy bands to signify the degree of degeneracy in the AFM-coupled helices.  We have set $N_\text{Fourier}=50$ in both \textit{x} and \textit{y}. \revision{In the experimentally reported and related cycloidal spin spiral in Ref. \cite{bode_nature_07}, a spiral with a unit cell period of approx. 25-26 sites was observed.}}
    \label{fig: AFM}
\end{figure}

The dependence of the \textit{p}-wave order on the exact period (even/odd) of the magnetic texture for short magnetic texture periods is a cause for concern regarding the robustness of the phenomenon in realistic helimagnets where the exact period of the magnetic texture can be difficult to control. The challenge in particular resides in the reliance on spin-rotation symmetry as a key symmetry operation for systems with odd unit cells, a symmetry vulnerable to spin-orbit coupling in realistic crystal environments. We now proceed to show how the presence of AFM order between helical chains introduces a new translation operation $\hat{\boldsymbol{\tau}}_{1/2}^y$ which preserves $\hat{\mathcal{T}}\hat{\boldsymbol{\tau}}_{1/2}$ symmetry also for odd unit cells. This enables a robust way to realize $p$-wave magnetism free of fine-tuning requirements, which is crucial for potential applications.

When the helical chains are stacked antiferromagnetically in the \textit{y}-direction, the helical chains and their time-reversed counterparts are always present due to the AFM order. As such, the existence of a translation operation $\hat{\boldsymbol{\tau}}_{1/2}$ is guaranteed, independent of the even/odd nature of the unit cell. This is shown in Fig. \ref{fig: AFM} (a) where we plot the $N_\text{u.c.}^x\times N_\text{u.c.}^y$ unit cell in the case of FM and AFM interchain order. In the case of FM order, the two rows of the unit cell are not related by time-reversal and as such, the only possible lattice translation mapping the lattice onto its time-reversed counterpart must be along \textit{x}. As previously discussed, this operation exists only for even unit cells, giving rise to the even-odd effect discussed above. When the two rows are antiferromagnetically ordered, they are intrinsically related by the time-reversal operation and this guarantees the existence of a translation vector along \textit{y} which, combined with time-reversal, leaves the system invariant.  Importantly, the proposed system of AFM-coupled helimagnetic chains is not experimentally unrealistic, and a closely related system is realized \cite{bode_nature_07} in manganese atoms on tungsten (110). \revision{While the system studied is a cycloidal spin spiral instead of a helical spin spiral as studied in this paper, the coplanar cycloid would still depict the transverse \textit{p}-wave polarization and as such still be a relevant system for exploring the behavior of the \textit{p}-wave order, particular in the presence of structurally induced spin-orbit coupling}. 

In Fig. \ref{fig: AFM} (b)-(c), we plot \revision{spin} polarization (P) and deviation $\Delta p$ for a set of helical chains with AFM interchain coupling as shown in the right panel of Fig. \ref{fig: AFM} (a). In this case, we observe the SOC to have a larger effect on the maximum and average \revision{spin} polarization, giving rise to a decrease relative the polarization without SOC. The effect of SOC on the polarization is larger than in the case of FM order. A key observation is the absence of the even-odd effect in $\Delta p$ and magnetization, which we attribute to the above mentioned existence of a $\hat{\boldsymbol{\tau}}_{1/2}$ translation operation also for odd unit cells. We also comment on the significant increase in bandstructure polarization observed in the absence of SOC. We attribute this to the presence of an additional period $\varphi_y$ in the \textit{y}-direction as introduced in the previously discussed continuum model. As the length of the unit cell increases, the presence of an AFM order in \textit{y} ensures the existence of a helical texture for momenta $\boldsymbol{k}=(k_x,k_y\neq0)$ and thus also the induced $\langle S_x\rangle$ polarization induced by said texture. We note, however, that this additional period likewise has to satisfy $\hat{\mathcal{T}}\hat{\boldsymbol{\tau}}_{1/2}$ on the lattice for the stability of the enhanced polarization to be ensured in the presence of Rashba SOC. Another reason that the $x$ \revision{spin} polarization is stable against unit cell length is that a non-zero $x$-polarization is permitted in the limit of infinite unit cell length, in contrast to the case of FM-coupled helices. In this limiting case, our helical structure becomes an A-type antiferromagnet polarized in $\pm y$. As the bands of this system are spin-degenerate, a nonzero $\langle s_{n,\boldsymbol{k}}^x \rangle$ is permitted, but inherently not useful due to the guaranteed existence of a state with the same momentum, but opposite \revision{spin} polarization. To quantify the extent to which our AFM-coupled helices are approaching this limit, we plot the averaged level splitting $S$ which is the splitting in energy within pairs of bands, averaged over momenta and band pairs, 
\begin{equation}
    S=\frac{1}{2N_\text{unit}^x}\frac{1}{(N_\text{fourier})^2}\sum_{n=1}^{2N_x}\sum_{\boldsymbol{k}} |E_{2n,\boldsymbol{k}}-E_{2n+1, \boldsymbol{k}}|
\end{equation}
in Fig. \ref{fig: AFM}(d). Note that there are $4N_\text{unit}^x$ bands in total, making up $2N_\text{unit}^x$ pairs. While the average level splitting remains non-zero within the ranges of unit cell sizes studied here, we note that the relationship between the average level splitting and the polarization of the band structure is going to be important for any practical application of the \textit{p}-wave polarization.
 
We briefly comment here on the possibility of an electric polarization induced by the inhomogeneous magnetization texture \cite{cheong_natmat_07, fiebig_natrevmat_16}. It is known that when magnetically frustrated systems form inhomogeneous textures, such as cycloidal magnetization, the inverse Dzyaloshinskii-Moriya interaction can induce an electric polarization. Phenomenologically, this may be written as follows \cite{mostovoy_prl_05} for a cubic/square lattice system: $\vecP \propto[\vecM (\nabla\cdot\vecM) - (\vecM\cdot\nabla)\vecM]$. The system in Fig. \ref{fig: FM} has a helical texture, which according to the above equation produces $\vecP=0$. $N$ copies of this system stacked in the $y$-direction also produces no \revision{electric} polarization, and is thus consistent with metallic behavior for the electrons in the plane. On the other hand, when the helical chains are stacked antiferromagnetically, as in Fig. \ref{fig: AFM}, one obtains a constant electric polarization that points out-of-plane $\vecP \parallel \hat{\boldsymbol{z}}$. However, as experimentally observed in \cite{fei_nature_18}, such an \revision{electric} polarization does not prohibit metallic behavior in the $xy$-plane.

\section{Conclusion}
We have computed the spin polarization of the electron bands of helimagnets in the presence of Rashba spin-orbit coupling and shown that the presence of \textit{p}-wave spin polarization is closely related to the period of the magnetic texture in the helical chains as well as the interchain order. In particular, we have shown that for helical chains with FM interchain order, an even-odd effect arises where the \textit{p}-wave symmetry is broken for odd unit cells associated with time-reversal symmetry breaking and nonzero magnetization. We attribute this to the absence of a lattice translation $\hat{\boldsymbol{\tau}}$ which maps the time-reversed lattice with odd unit cells onto itself. We resolve this strong dependence on unit cell length by introducing an antiferromagnetic coupling between the helical chains, thus explicitly enforcing a time-reversal relation between adjacent chains. This helical order closely corresponds to experimentally realizable systems such as Mn atoms on tungsten (110) and we found that it guarantees the \textit{p}-wave symmetry for all unit cell lengths. These results provide a route to provide robust $p$-wave spin polarization regardless of the presence of SOC and regardless of the precise period of the helimagnetic spiral.

\acknowledgments

We thank P. Sukhachov, S. M. Selbach, B. Brekke, H. G. Giil, and A. Brataas for useful discussions. This work was supported by the Research
Council of Norway through Grant No. 323766 and its Centres
of Excellence funding scheme Grant No. 262633 “QuSpin.” Support from
Sigma2 - the National Infrastructure for High Performance
Computing and Data Storage in Norway, project NN9577K, is acknowledged.

\bibliography{references}

\end{document}